\documentclass[aip,pof]{revtex4-1} %aip,pof

\usepackage{graphicx}%
\usepackage{graphics}%
\usepackage{xspace}%

\newcommand{\micron}{$\mu$m\xspace}     % micrometer
\newcommand{\us}{$\mu$s\xspace}         % microsecond
\newcommand{\Celsius}{$^\circ$C\xspace} % degrees Celsius

\begin{document}

\title{Breakup of diminutive {Rayleigh} jets}

\author{Wim van Hoeve}
\author{Stephan Gekle}
\author{Jacco H.~Snoeijer}
\author{Michel Versluis} \affiliation{Physics of Fluids, Faculty of Science and Technology, and MESA$^+$ Institute for Nanotechnology, University of Twente, P.O.~Box 217, 7500 AE Enschede, The Netherlands}
\author{Michael P.~Brenner} \affiliation{School of Engineering and Applied Sciences, Harvard University, Cambridge, Massachusetts 02138, USA}
\author{Detlef Lohse} \affiliation{Physics of Fluids, Faculty of Science and Technology, and MESA$^+$ Institute for Nanotechnology, University of Twente, P.O.~Box 217, 7500 AE Enschede, The Netherlands}



\date{\today}

\begin{abstract}
Discharging a liquid from a nozzle at sufficient large velocity leads to a continuous jet that due to capillary forces breaks up into droplets. Here we investigate the formation of microdroplets from the breakup of micron-sized jets with ultra high-speed imaging. The diminutive size of the jet implies a fast breakup time scale $\tau_\mathrm{c} = \sqrt{\rho r^3 / \gamma}$ of the order of 100\,ns{}, and requires imaging at 14 million frames per second. We directly compare these experiments with a numerical lubrication approximation model that incorporates inertia, surface tension, and viscosity [Eggers and Dupont, J.~Fluid Mech.~\textbf{262}, 205 (1994); Shi, Brenner, and Nagel, Science, \textbf{265}, 219 (1994)]. The lubrication model allows to efficiently explore the parameter space to investigate the effect of jet velocity and liquid viscosity on the formation of satellite droplets. In the phase diagram we identify regions where the formation of satellite droplets is suppressed. We compare the shape of the droplet at pinch-off between the lubrication approximation model and a boundary integral (BI) calculation, showing deviations at the final moment of the pinch-off.
Inspite of this discrepancy, the results on pinch-off times and droplet and satellite droplet velocity obtained from the lubrication approximation agree with the high-speed imaging results.
\end{abstract}

\maketitle

\section{Introduction}
\begin{figure}[t!]
    \includegraphics[]{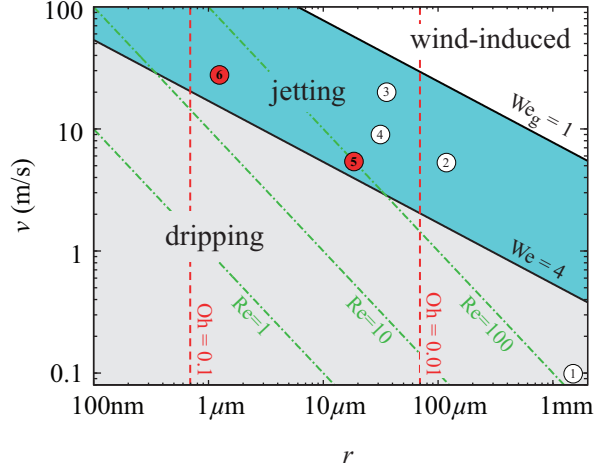}
    \caption{\label{Fig:breakup_regimes} Classification of droplet formation regimes for a liquid discharging an orifice of radius $r$ with liquid velocity $v$. Droplet formation through the Rayleigh breakup mechanism in jetting (blue) is bounded by a lower and upper critical jet velocity, expressed in terms of the Weber number $\textrm{We} = \rho_\ell v^2 r / \gamma > 4$ and $\textrm{We}_\mathrm{g} = (\rho_\mathrm{g} / \rho_\ell) \textrm{We} < 1$ respectively. The lines shown correspond to the Weber number, Reynolds number ($\textrm{Re} = \rho_\ell v r / \eta$), and Ohnesorge number ($\textrm{Oh}=\eta / \sqrt{\rho_\ell r \gamma}$) for pure water (with density $\rho_\ell = 1000$\,kg/m$^3$, surface tension $\gamma = 72$\,mN/m, and viscosity $\eta = 1\,\mbox{mPa}\cdot$s). The encircled numbers (1--4) refer to droplet formation studies performed by Ambravaneswaran \emph{et al.},\cite{Ambravaneswaran2000} Kalaaji \emph{et al.},\cite{Kalaaji2003} Gonz\'alez \& Garc\'ia,\cite{Gonzalez2009} and Pimbley \& Lee\cite{Pimbley1977} respectively; in this work we study diminutive Rayleigh jets using ultra high-speed imaging at 500\,kfps (5) and 14\,Mfps (6).}
\end{figure}
A narrow size distribution in droplet formation is important in many industrial and medical applications.
For example,
the controlled formation of (double)emulsions from microfluidic devices in T-shaped\cite{Thorsen2001, vanSteijn2010} or flow-focusing geometries\cite{Anna2003, Utada2005, Castro2009} is needed in many personal care products, foods, and cosmetics.
In food industry, the production of powders with a monodisperse particle size distribution through spray-drying results in a reduction of transportation and energy costs.
In inkjet printing, monodisperse microdroplets are required to accurately control droplet deposition.\cite{Bogy1979}
In drug inhalation technology, monodisperse droplets lead to an improved lung targeting.\cite{Mitchell1987, Bennett2002}

Highly monodisperse micrometer-sized droplet production is achieved in various droplet generators.\cite{Basaran2002, Basaran2007} In continuous jet technology the breakup of a liquid jet emanating from a nozzle is stimulated by an acoustic wave, resulting in a continuous stream of droplets.\cite{Le1998} Piezoelectric drop-on-demand systems generate a well-defined piezo-driven pressure pulse into the (ink)reservoir which forces a precisely controlled amount of liquid to detach from the nozzle.\cite{Xu2007, Wijshoff2010}
In electrospray atomization a high voltage is applied to generate (sub)micron-sized droplets of conducting liquids.\cite{Tang1994, Chen1995}
However, for drug inhalation technology ideally one would need a low-cost droplet generator that is simple, light, and disposable.
The typical size of nebulizer droplets is 2.5\,\micron in radius.
Droplets that are too large do not penetrate into the deeper regions of the lungs, whereas droplets that are too small evaporate or are exhaled.\cite{Mitchell1987, Bennett2002} Inhaler sprays therefore need a well-controlled and narrow size distribution.
The more sophisticated droplet generators involve the use of dedicated equipment and hence do not meet the criteria for this specific application.

The formation of droplets by the slow emission of a liquid from a nozzle (\emph{e.g.}~a leaking faucet) forms a pendant droplet that grows slowly, characterized by a quasi-static balance between inertial and surface tension forces.\cite{Brenner1997, Ambravaneswaran2000, Ambravaneswaran2004, Coullet2005} The droplet formation mechanism in this regime is associated as `dripping'.
It is known that the shape of the nozzle opening can dramatically influence the size of the droplets.\cite{Chen2004} Droplet formation in the dripping regime typically produces large droplets at low production rates.

When the liquid flow rate is progressively increased such that the liquid velocity $v$ is sufficiently large such that the kinetic energy overcomes the surface energy a continuous liquid jet is formed.
The lower critical velocity for jet formation can be expressed in terms of the Weber number
\begin{equation}
\label{Eq:Weber}
    \textrm{We} = \frac{\rho_\ell r v^2}{\gamma} > 4,
\end{equation}
with radius of the jet $r$, liquid density $\rho_\ell$, and surface tension $\gamma$.\cite{Lin1998}
The formation of a jet -- that is inherently unstable -- gives rise to the next droplet formation regime, where droplets are generated by the spontaneous breakup of the jet to minimize its surface energy. Droplet formation by `jetting' a liquid is referred to as `Rayleigh breakup' as described by Plateau\cite{Plateau1873} and Lord Rayleigh\cite{Rayleigh1879} more than a century ago. A small disturbance introduced by mechanical vibrations or by thermal fluctuations will grow when its wavelength exceeds the circumference of the jet. The optimum wavelength for an inviscid liquid jet is expressed as $\lambda_\mathrm{opt} = 2\sqrt{2} \pi r$ and is determined by the jet radius only.\cite{Rayleigh1879} The system automatically selects this optimum wavelength and breaks up in fixed fragments of volume $\lambda_\textrm{opt} \pi r^2$, which then determines the droplet size. The size of the droplets is thus governed by the geometry of the system and it is independent on the jetting velocity.

When the liquid velocity is further increased the relative velocity between the jet and the ambient air can no longer be neglected. Aerodynamic effects accelerate the breakup process and a shortening of the length from the nozzle exit to the location of droplet pinch-off is observed. A transition from the `Rayleigh breakup' regime to the first wind-induced breakup regime occurs when the inertia force of the surrounding air reaches a significant fraction of the surface tension force, such that the Weber number in gas
\begin{equation}
\label{Eq:Weber_gas}
    \mathrm{We}_\mathrm{g} = \frac{\rho_g}{\rho_\ell} \mathrm{We} > 0.2,
\end{equation}
with $\rho_\mathrm{g}$ the density of the gas.\cite{Ranz1956}

Figure \ref{Fig:breakup_regimes} shows a classification of droplet formation regimes based on the radius of the jet $r$ and the liquid velocity $v$. All the lines in the figure are indicated for pure water. Droplet formation through the breakup of a continuous liquid jet in the Rayleigh breakup regime (`jetting') is bounded by a lower and upper critical velocity of 5\,m/s and 35\,m/s respectively for a 10\,\micron radius jet, and 17\,m/s and 110\,m/s for a 1\,\micron jet. From a practical prospective, the velocity operating range for Rayleigh breakup becomes wider for decreasing jet size. The size of the droplets is governed by the jet radius and is independent on the jetting velocity. This makes it the favorable regime for high-throughput monodisperse microdroplet formation.
A complete overview on liquid jet breakup has recently been given by Eggers and Villermaux.\cite{Eggers2008}

Droplet formation by dripping from the tip of a nozzle or through the breakup of a continuous liquid jet has been extensively studied both experimentally and numerically. In the work of Wilkes \emph{et al.}\cite{Wilkes1999}~a two-dimensional finite element method (FEM) is used to study the formation of droplets and a comparison is made to high-speed imaging results recorded at 12,000 frames per second (fps). Hilbing \emph{et al.}\cite{Hilbing1996}~used a non-linear boundary element method (BEM) to study the effect of large amplitude periodic disturbances on the breakup of a liquid jet into droplets. In the work of Moseler and Landman a molecular dynamics (MD) simulations for nanometer-sized jets was presented.\cite{Moseler2000}

Fully three-dimensional or axisymmteric analysis of free-surface flows in droplet formation represents a complicated and computationally intensive task. The use of one-dimensional (1D) simplified models based on the lubrication approximation became standard since the work of Eggers and Dupont,\cite{Eggers1994} Shi, Brenner and Nagel,\cite{Shi1994} and Brenner \emph{et al.}\cite{Brenner1997} Application of the lubrication approximation offers a fast computational algorithm for describing free-surface flows, while incorporating the effect of inertia, surface tension, and viscosity. Many groups have successfully implemented the lubrication approximation to study the dynamics of droplet formation (in both `dripping' and `jetting') and the predictions are remarkable accurate in comparison to results obtained experimentally or by more complex numerical simulations.
In Ambravaneswaran \emph{et al.}\cite{Ambravaneswaran2002}~the performance of a one-dimensional model based on the lubrication approximation is evaluated by comparing it with the predictions obtained from a two-dimensional FEM calculations.
Ambravaneswaran \emph{et al.}\cite{Ambravaneswaran2000}~studied the formation of a sequence of hundreds of droplets by dripping from a nozzle using a one-dimensional model based on the lubrication approximation and compared the rich non-linear dynamics with experimental results.
In Yildirim \emph{et al.}\cite{Yildirim2005}~it is stated that results obtained from a one-dimensional model can be used to improve the accuracy of the \emph{drop weight method} to accurately measure the surface tension of a liquid.
In Furlani \emph{et al.}\cite{Furlani2010}~a one-dimensional analysis of microjet breakup is studied within the lubrication approximation and validated using volume of fluid simulations.

Relatively little attention has been given to the experimental validation of such models in predicting the formation of microdroplets from the breakup of microjets. A study on the dynamics of these microscopically thin jets is experimentally extremely difficult due to the small length and time scales involved.\cite{Basaran2002}
The local thinning of the liquid microjet followed by the droplet pinch-off is an extremely fast process. If the liquid viscosity can be safely neglected -- \emph{i.e.} when the Ohnesorge number
\begin{equation}
    \label{Eq:Oh}
    \textrm{Oh} = \frac{\mu}{\sqrt{\gamma \rho_\ell r}} \ll 1,
\end{equation}
with liquid viscosity $\mu$ -- the collapse is driven by a balance between inertial and surface tension forces. The relevant time scale is then given by the capillary time
\begin{equation}
    \tau_\mathrm{c} = \sqrt{\frac{\rho_\ell r^3}{\gamma}}.
\end{equation}
In the specific case for the breakup of a 1\,\micron liquid jet, the capillary time is of the order of 100\,nanoseconds.

In this paper we study microdroplet formation from the breakup of a continuous liquid microjet experimentally using ultra high-speed imaging and within a 1D lubrication approximation model. We focus on two distinct cases for different jet sizes. First, we resolve the breakup of a 18.5\,\micron jet (\emph{cf.}~number 5 in Fig.~\ref{Fig:breakup_regimes}) at a high spatial and temporal resolution, such that the smallest structures (\emph{i.e.}~satellite droplets) can be studied in great detail. We make a direct comparison between the high-speed imaging results captured at 500\,kfps and those obtained from the 1D model calculation for this specific case. One of the key contributions of this paper is to demonstrate the good agreement between the experimental and lubrication approximation results, which to date have never been established before on such a small time and length scale (\emph{cf.}~the encircled numbers 1--4 indicating existing studies presented in literature in Fig.~\ref{Fig:breakup_regimes}). For the second case we push towards the experimental limit (number 6 in the same figure) and provide ultra high-speed imaging results recorded at 14\,Mfps (corresponding to 73\,ns interframe time) of microdroplet formation from the breakup of a 1\,\micron diminutive jet.

The paper is organized as follows. In section \ref{Sec:lubrication_approximation} we briefly describe the one-dimensional model based on the lubrication approximation. The experimental setup to resolve the extremely fast droplet formation process is described in section \ref{Sec:setup}, while the results are presented in section \ref{Sec:results}. Section \ref{Sec:setup} and section \ref{Sec:results} are divided into two sub-sections for both cases of jet breakup: for a microjet (A) and for a diminutive jet of 1\,\micron in size (B). In section \ref{Sec:bi} a comparison is made between the 1D model and a boundary integral calculation. A discussion and concluding remarks are given in section \ref{Sec:discussion_conclusion}.

\section{\label{Sec:lubrication_approximation}Lubrication approximation}
\begin{figure}
    \includegraphics[]{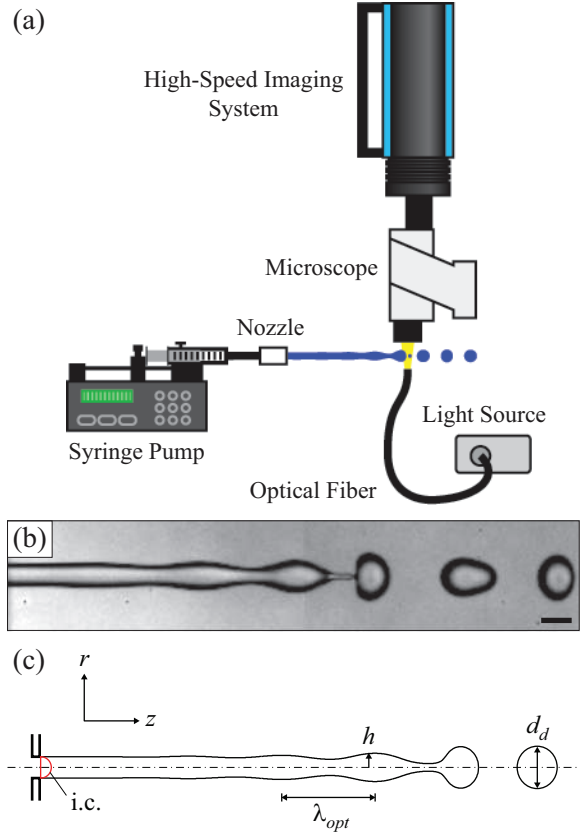}
    \caption{\label{Fig:setup}(a) Schematic overview of the experimental setup to visualize the extremely fast breakup of a liquid microjet into droplets. (b) Snapshot of a high-speed imaging recording of the breakup of a 18.5\,\micron radius jet. The scale bar denotes 50\,\micron. (c) Coordinate system. The initial condition (i.c.)~of the system is indicated by the solid red line.}
\end{figure}

Numerical simulations of jet breakup are carried out by solving the Navier-Stokes equations within the lubrication approximation as described in detail in the work by Eggers and Dupont\cite{Eggers1994} and Shi, Brenner and Nagel.\cite{Shi1994} It was shown that droplet formation -- \emph{e.g.}~from a dripping faucet -- can be predicted within great accuracy by solving the Navier-Stokes equation in a one-dimensional approximation obtained from a long-wavelength expansion.
We apply the very same technique to model microdroplet formation from the breakup of an axisymmetric liquid microjet emanating from a circular orifice.
The geometry of the system is schematically shown in Fig.~\ref{Fig:setup}(c). For small perturbations of the jet, the radial length scale is very much smaller than the longitudinal length scale and it is therefore appropriate to apply the lubrication approximation. We note, however, that close to break-up this approximation is no longer justified. One of the key questions is to investigate whether this influences the prediction of the breakup phenomenon.

Here we briefly review the main steps leading to the lubrication equations. The state of the system is described by the pressure field $p$ and the velocity field $v$. A Taylor series expansion around $r = 0$ is obtained for both the velocity component in the $z$-direction $v_z$ and the pressure field $p$.
The velocity component in the $z$-direction is represented by a uniform base flow $v_0$ with a second-order correction term. The velocity in the radial direction $v_r$ follows from continuity.
The linearized equations are inserted in the full Navier-Stokes equation in cylindrical form and keeping only the lowest order in $r$.

The system is closed by applying the boundary conditions for the normal and tangential force on the jet surface. The normal force is balanced by the Laplace pressure, which gives the pressure jump across the interface. For the thin jet of our experiment at moderate jetting velocity, the breakup is not influenced by the surrounding air and the tangential force is set to zero.
Applying these boundary conditions gives the reduced form of the Navier-Stokes equation
\begin{equation}
\label{Eq:RedNS}
    \frac{\partial v}{\partial t} + v v^\prime = - \frac{\gamma}{\rho_\ell} C^\prime + \frac{3 \mu}{\rho_\ell} \frac{\left(h^2 v^\prime\right)^\prime}{h^2},
\end{equation}
where $h$ is the radius of the jet, $v$ is the liquid velocity, and prime denotes the derivative with respect to the axial-coordinate $z$. The curvature of the interface $C$ is given by
\begin{equation}
\label{Eq:curvature}
     C = \frac{1}{R_1} + \frac{1}{R_2} = \frac{1}{h \sqrt{1 + {h^\prime}^2}} - \frac{h^{\prime\prime}}{\left(1 + {h^\prime}^2\right)^{3 / 2}}.
\end{equation}
The interface moves with the velocity field as
\begin{equation}
\label{Eq:Interface}
    \frac{\partial h^2}{\partial t} + \left( v h^2 \right)^\prime = 0.
\end{equation}

The set of linear equations (Eq.~\ref{Eq:RedNS}, \ref{Eq:curvature}, and \ref{Eq:Interface}) is solved using an explicit scheme ODE solver in MATLAB (The Mathworks Inc., Natick, MA, USA).\cite{note1} A fixed number of grid points are homogeneously distributed from the nozzle exit to the tip of the jet. %Note that the flow inside the nozzle is not included in the computational domain.

As the initial condition for the shape of the jet, we used a hemispherical droplet described by $h = \sqrt{ h_0^2 - z^2 }$ (see Fig.~\ref{Fig:setup}(c)), with $h_0$ the initial radius of the jet at the nozzle exit, and axial coordinate $z$ containing 1000 grid points, homogeneously distributed between the nozzle exit (at $z = 0$) and the tip of the jet at $z = h_0$.
The initial jetting velocity $v_\mathrm{jet}$ was constant along the $z$-axis.
A modulation of the nozzle radius is applied in the numerical simulations to initiate jet breakup, mimicking thermal fluctuations. The amplitude of variation of the jet radius at the nozzle exit is small, $\delta / h_0 \approx 0.005$.
\begin{equation}
    h_\mathrm{nozzle} = h_0 + \delta{} \sin{ 2 \pi{} f  t }
\end{equation}
with $f$ the driving frequency.
To assure a constant flow rate $Q$ through the nozzle, the velocity was modulated correspondingly,
\begin{equation}
    v_\mathrm{nozzle}(t) = \frac{h_0^2 v_\mathrm{jet}(t)}{h_\mathrm{nozzle}^2}.
\end{equation}

The driving frequency of the modulation is chosen such that it matches the optimum wavelength for jet breakup ($f = v_\mathrm{jet} / \lambda_\mathrm{opt}$).
The amplitude of the wave grows until it equals the radius of the jet and a droplet pinch-off occurs. The moment of droplet pinch-off is defined as the moment where the minimum width of the jet is below a threshold value, which in our simulations we set to 100\,nm. After pinch-off all grid points between the nozzle exit and the pinch-off location were redistributed and the calculations continued until another droplet was formed.
After droplet pinch-off, the continuous jet and the detached droplet where calculated individually. When a droplet meets another droplet (or jet) the two objects where combined to simulate droplet coalescence. Droplet coalescence is defined when two objects overlap with a critical distance $0.005 h_0$.

\section{\label{Sec:setup}Experimental setup}
The experimental study of microjet breakup is extremely challenging due to the small length and time scales involved. This section describes the experimental setup that is used to visualize the breakup of liquid microjets. We consider two different systems -- first for the breakup of a 18.5\,\micron liquid microjet, and second, for a diminutive jet under the extreme condition for resolving its breakup using visible light microscopy.

\subsection{\label{SubSec:microjets}Microjets (18.5\,\micron)}
Fig.~\ref{Fig:setup}(a) shows the experimental setup to visualize the formation of microdroplets from the breakup of a continuous liquid microjet.
An aqueous solution of 40\% w/w glycerol, dissolved in a 0.9\% w/w saline solution (with $\rho_\ell = 1098$\,kg/m$^3$, $\eta = 3.65\,\mbox{mPa}\cdot$s, and $\gamma = 67.9$\,mN/m), was supplied at a constant flow rate $Q$ of 0.35\,ml/min through a high-precision syringe pump (accuracy 0.35\%) (PHD 22/2000, Harvard Apparatus, Holliston, MA).
The addition of sodium chloride prevents the droplets from charging and avoids deflection of the jet when it exits the nozzle. The liquid is forced to flow through a silicon micro machined nozzle chip (\mbox{Medspray} XMEMS bv, The Netherlands).
The nozzle chip consist of a rectangular opening of aspect ratio 8:1 ($W \times H = 89\,\mbox{\micron} \times 11\,\mbox{\micron}$). This leads to a jet of non-circular cross-section, with the major/minor axis switching along the jet.\cite{Rayleigh1879, Gutmark1999} The oscillation is damped out by viscosity and once the jet gets back to a cylindrical shape its radius is measured to be $r = 18.5$\,\micron. The jet velocity $v_\mathrm{jet}$ is calculated from the imposed liquid flow rate and the cross-sectional area of the jet $A$ as $v_\mathrm{jet} = Q / A = 5.4$\,m/s, with $A = \pi r^2$.
The experimental conditions are expressed by the dimensionless Reynolds number ($\textrm{Re}$), Weber number for the liquid ($\textrm{We}$), Weber number for the ambient gas ($\textrm{We}_\mathrm{g}$), and the Ohnesorge number ($\textrm{Oh}$). The Reynolds number $\textrm{Re} = \sqrt{\textrm{We}} / \textrm{Oh} \approx 30$, assuring that the flow remains laminar. The Weber number for the liquid and the gas, defined in Eq.~\ref{Eq:Weber} and Eq.~\ref{Eq:Weber_gas}, are respectively $\textrm{We} \approx 8.7$ and $\textrm{We}_\mathrm{g} \approx 0.01$, confirming that the breakup of the liquid jet is purely driven by the Rayleigh breakup mechanism, since the criteria $\textrm{We} > 4$ and $\textrm{We}_\mathrm{g} < 1$ holds.\cite{Lin1998} The Ohnesorge number, defined in Eq.~\ref{Eq:Oh}, $\textrm{Oh} \approx 0.10$ is small, but finite, which implies that the viscosity of the liquid influences the motion of the fluid. The relevant time scale for the motion of the liquid is given by the capillary time $\tau_\mathrm{c} = \sqrt{\rho_\ell r^3 / \gamma} \approx 10$\,\us.
To resolve the droplet formation and the extremely fast pinch-off, a high-speed camera (Hypervision HPV-1, Shimadzu Corp., Kyoto, Japan) was used which captures 102 consecutive images at a frame rate of 500,000 frames per second (fps) and $312\,\times\,260$ pixel spatial resolution, \emph{cf.}~Fig.~\ref{Fig:setup}(b).
To minimize motion blur the exposure time of the camera was set to 1\,\us.
The camera was mounted to a microscope (BX-FM, Olympus Nederland bv, Zoeterwoude, The Netherlands) with a ${50\times}$ objective lens (SLMPlan N ${{50\times}/0.35}$, Olympus). The system was operated in bright-field mode using a high-intensity continuous light source (LS-M352A, Sumita Optical Glass Europe GmbH, Germany) and fiber illumination.

\subsection{\label{SubSec:diminutive_microjets}Diminutive microjets (1\,\micron)}
Diminutive liquid microjets were studied by mounting a nozzle chip that consist of 49 orifices of radius $r = 1.25$\,\micron separated from each other by 25\,\micron in direction along a row. A 0.9\% saline solution (with $\rho_\ell = 1003$\,kg/m$^3$, $\eta = 1\,\mbox{mPa}\cdot$s, and $\gamma = 72$\,mN/m) was discharged at a constant liquid flow rate of 0.50\,ml/min from the nozzle chip resulting in the formation of 49 parallel microjets of approximately 1.25\,\micron in radius (neglecting the \emph{vena contracta} effect). The jet velocity is estimated by $v_\mathrm{jet} = Q / (n A) = 35$\,m/s, with $n = 49$ the number of jets, and $A = \pi r^2$ the cross-sectional area of a single orifice. The Reynolds number $\textrm{Re} = 44$; the Weber number for the liquid $\textrm{We} = 21$, and for the gas $\textrm{We}_\mathrm{g} = 0.03$; the Ohnesorge number $\textrm{Oh} = 0.10$. The capillary time is $\tau_\mathrm{c} = 165$\,ns.
The extremely fast dynamics involved in the breakup of these diminutive jets were captured with the ultra high-speed Brandaris 128 camera\cite{Chin2003} at a framerate of 13.76\,Mfps, corresponding to a temporal resolution of 73\,ns. The microscopic system used was the same as described above. A high-intensity flash lamp coupled into a liquid lightguide was used that produces a single flash with a pulse duration sufficiently long to expose all 128 image frames of the high-speed recording ($\approx 10$\,\us).

\begin{figure*}
    \includegraphics[]{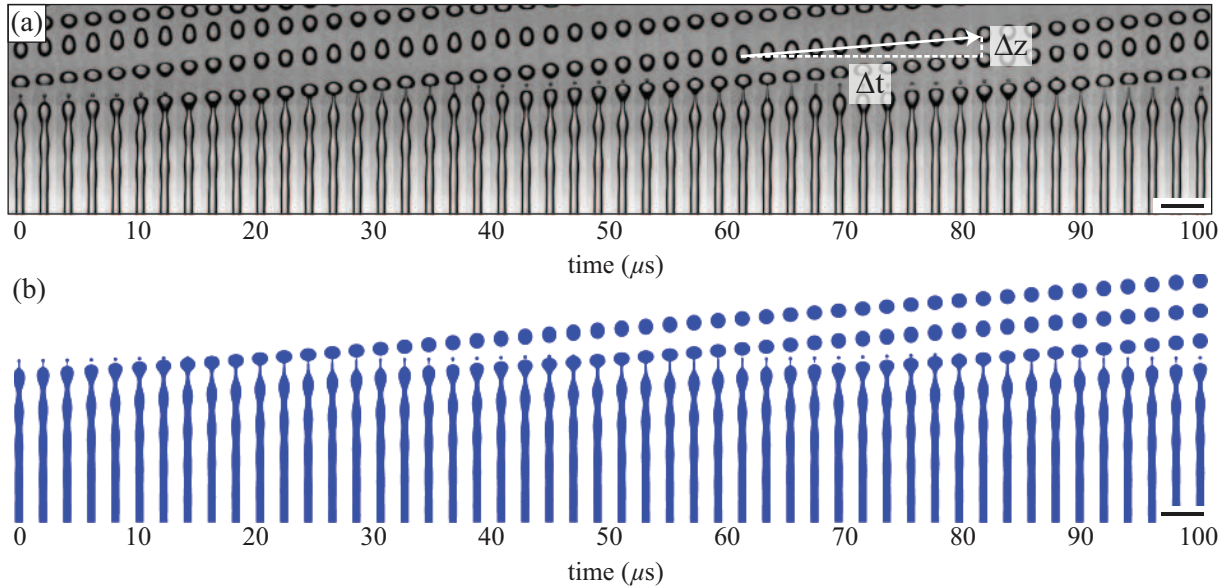}
    \caption{\label{Fig:timeseries}Comparison between the time series of the high-speed imaging results (a) and those obtained from the lubrication approximation calculation (b), indicating that droplet formation is predicted in good agreement with the experimental results. The droplet velocity is calculated from the displacement of the droplet's center of mass between two consecutive images. The scale bar in both panels correspond to 200\,\micron.}
\end{figure*}
\section{\label{Sec:results}Results}
\subsection{Results for microjets}
In Fig.~\ref{Fig:timeseries} we show a time series of the formation of a droplet captured using ultra high-speed imaging at 500\,kfps (a) and we show the accompanying prediction based on the lubrication approximation model (b). The lubrication approximation model calculation is based on the experimental values for the radius of the jet $r = 18.5$\,\micron, the jet velocity $v_\mathrm{jet} = 5.4$\,m/s, and the liquid properties only. We observe two different types of droplets --- the primary droplet, with a size almost twice the diameter of the jet, and a small satellite droplet. The satellite droplet is formed from the breakup of the thin thread between the jet and the primary droplet.
In Fig.~\ref{Fig:timeseries} it is demonstrated that the lubrication model accurately predicts both the formation of the primary droplet and its satellite droplet.

\begin{figure}
    \includegraphics[]{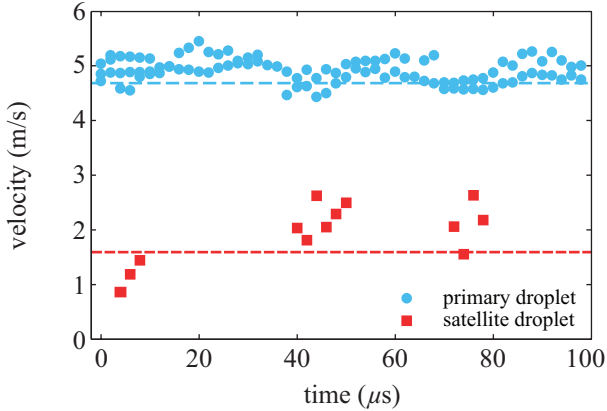}
    \caption{\label{Fig:velocity}Experimental droplet velocity of the primary droplet (blue) and the satellite droplet (red) as a function of the time. The dashed line is the velocity predicted by the lubrication approximation model. }
\end{figure}
We obtain the velocity of the droplets from the displacement of the droplet's center of mass between two frames. Eight droplets are traced throughout the recording, resulting in a total of 116 velocity measurements, which are shown in Fig.~\ref{Fig:velocity}.
The predicted velocities for the primary droplet and the satellite droplets of respectively $4.68\,\pm\,0.01$\,m/s and $1.6\,\pm\,0.3$\,m/s, nicely agree with the experimental findings of $4.9\pm{}0.2$\,m/s and $1.9\pm{}0.6$\,m/s. In Fig.~\ref{Fig:velocity} it is also displayed that the velocity of the primary droplet shows a minimum (at time 6\,\us, 40\,\us, and 70\,\us) that is correlated with the existence of the satellite droplet.
This periodicity in the primary droplet velocity is due to an inertia effect as a consequence of the pinch-off.

\begin{figure*}
    \includegraphics[width=16cm]{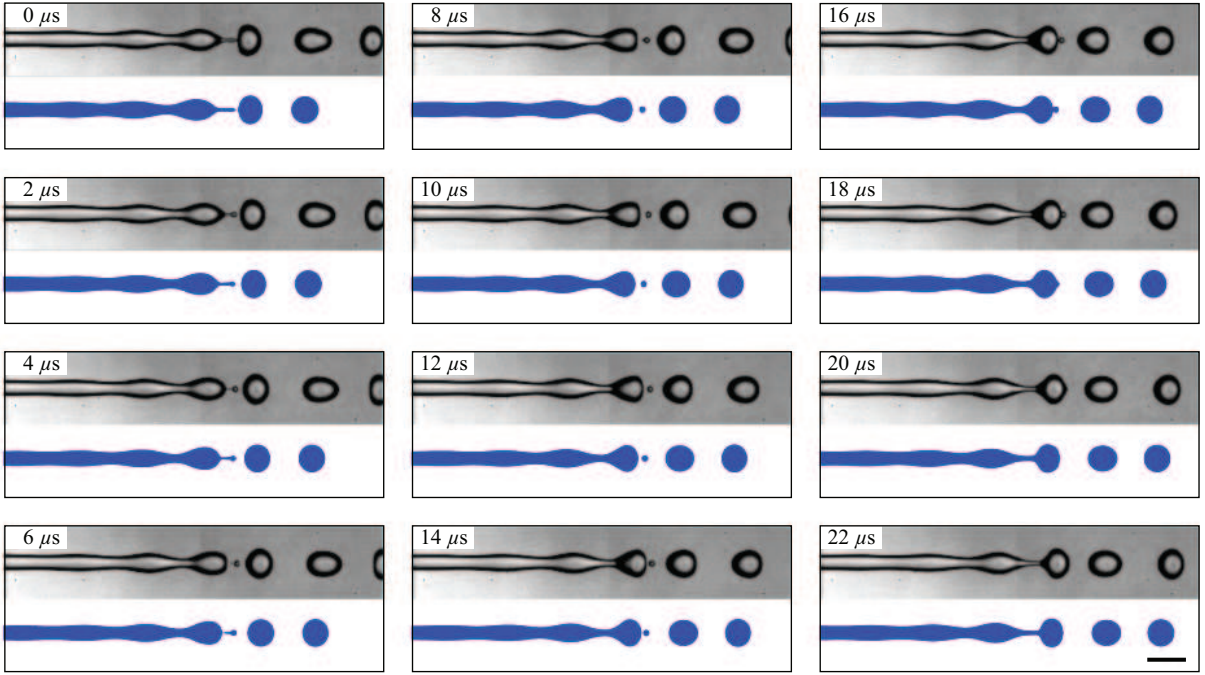}
    \caption{\label{Fig:detailed_comparison}Direct comparison between experimental results and the lubrication approximation model in both space and time. High-speed imaging at 500,000\,fps shows the breakup of a 18.5\,\micron radius jet into a primary droplet and a small satellite droplet. A 40\% glycerol-saline solution is supplied at 0.35\,ml/min, equivalent to a jetting velocity of 5.4\,m/s. The lubrication calculation (in blue) is synchronized to the location of droplet pinch-off in the experiment at $t = 0$\,\us. No fitting parameters were used. Satellite droplet merging with the primary droplet was not included in the model. The scale bar in the lower right corner denotes 100\,\micron.}
\end{figure*}
The primary droplet travels at a speed that is smaller than the imposed liquid velocity $v_\mathrm{d} < v_\mathrm{jet}$, which is caused by the loss of kinetic energy due to surface oscillations after droplet pinch-off. The reason for the deceleration of the satellite droplet with respect to the mean jet velocity can be understood from Fig.~\ref{Fig:detailed_comparison}. In this figure we make a detailed comparison between the experimentally measured shape of the jet and the prediction obtained from the lubrication approximation model. The moment of primary droplet pinch-off defines time $t = 0$\,\us in the top-left panel. The axial-coordinate of the pinch-off location in the high-speed imaging recording is aligned with the one in the model calculation (in the first frame only).
The satellite droplet pinches off from the primary droplet (at $t \approx 0$\,\us) before it pinches off from the jet (at $t \approx 4$\,\us), resulting in a deceleration of the satellite droplet.\cite{Pimbley1977}

\begin{figure}
    \includegraphics[]{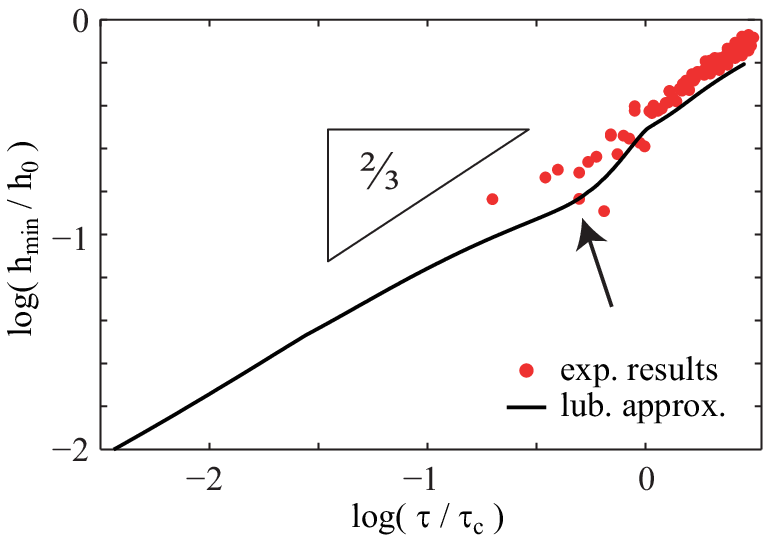}
    \caption{\label{Fig:scaling}The logarithm of the normalized minimum radius of the neck $h_\mathrm{min} / h_0$ as a function of the logarithm of the time remaining until droplet pinch-off $\tau = t_\mathrm{c} - t$, normalized by the capillary time $\tau_\mathrm{c}$, scales with a characteristic 2/3 scaling exponent.}
\end{figure}
In Fig.~\ref{Fig:scaling} we plot the evolution of the minimum radius of the neck $h_\mathrm{min}$ during collapse, until breakup.
When the viscosity of the liquid is neglected, the collapse of the `neck' can be described by a radially collapsing cylinder. From a balance between inertia and surface tension forces it follows that the minimum radius of the neck $h_\mathrm{min} \propto (\tau / \tau_\mathrm{c})^{2/3}$, with a characteristic 2/3 exponent.\cite{Leppinen2003}
The experimental data fall on a single power-law curve with an exponent equal to 2/3.
\begin{figure}
    \includegraphics[]{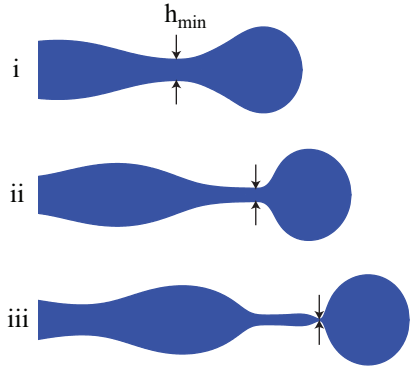}
    \caption{\label{Fig:snaps}The shape of the neck for three successive moments approaching droplet pinch-off.}
\end{figure}
The evolution of the minimum radius of the neck during pinch-off predicted by the lubrication approximation shows a deviation from the $2/3$ slope line (indicated by the arrow in Fig.~\ref{Fig:scaling}), which can be attributed to the formation of the satellite droplet, as described in the work by Brenner \emph{et al.}~\cite{Brenner1997} and Notz \emph{et al.}\cite{Notz2001}
This is illustrated in Fig.~\ref{Fig:snaps} where three moments during droplet pinch-off are shown. In (i) and (ii) the thin liquid thread shows an elongation, which is followed up by a thickening (iii) and the formation at a satellite droplet. This affects the breakup dynamics and causes a (transient) deviation from the asymptotic 2/3 behavior.

\subsection{Results for diminutive microjets}
\begin{figure*}
    \includegraphics[]{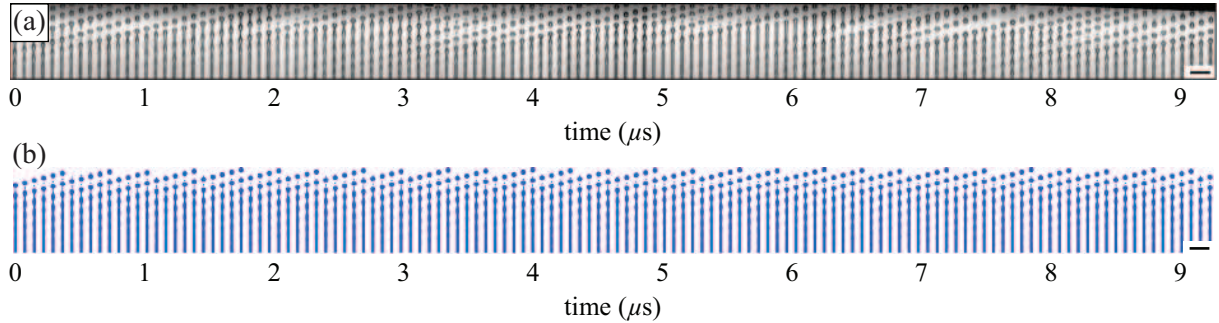}
    \caption{\label{Fig:timeseries_brandaris} Time series of the ultra high-speed imaging results recorded at 13.76\,Mfps (a) and the accompanying prediction from the lubrication approximation model (b) showing the breakup of a 1.25\,\micron liquid jet into droplets. The period in droplet formation is approximately 300\,ns. The interframe time is 73\,ns. The high-speed image frames are displayed in greater detail in Fig.~\ref{Fig:montage}. The scale bar in the lower right corner of both panels denote 20\,\micron.}
\end{figure*}
We now consider the results obtained from the ultra high-speed imaging recordings for the breakup of diminutive jets of 1\,\micron in radius. Fig.~\ref{Fig:timeseries_brandaris} shows a time series obtained from the 128 image frames captured with the Brandaris 128 camera\cite{Chin2003} (a) and the accompanying lubrication approximation calculation (b). The experimental results show the formation of microdroplets at a temporal resolution of 73\,ns and a spatial resolution of 0.22\,\micron/pixel. Note that the total time displayed is less then 10\,\us{} -- within this time 30 droplets and 30 satellite droplets are formed. The ultra high-speed time series shows the formation of the primary droplets and even of the existence of satellite droplets.
The primary droplet size is determined by measuring the cross-sectional area of the droplet using digital image analysis -- its equivalent radius is $2.5\,\pm\,0.2$\,\micron, which nicely agrees with the predicted droplet size by the lubrication model of 2.4\,\micron. The area of the satellite droplet is too small to measure it accurately. From the lubrication approximation calculation we can learn that the size of the satellite droplet is 0.6\,\micron (equivalent to a droplet volume of 1\,fl).

\begin{figure*}
    \includegraphics[]{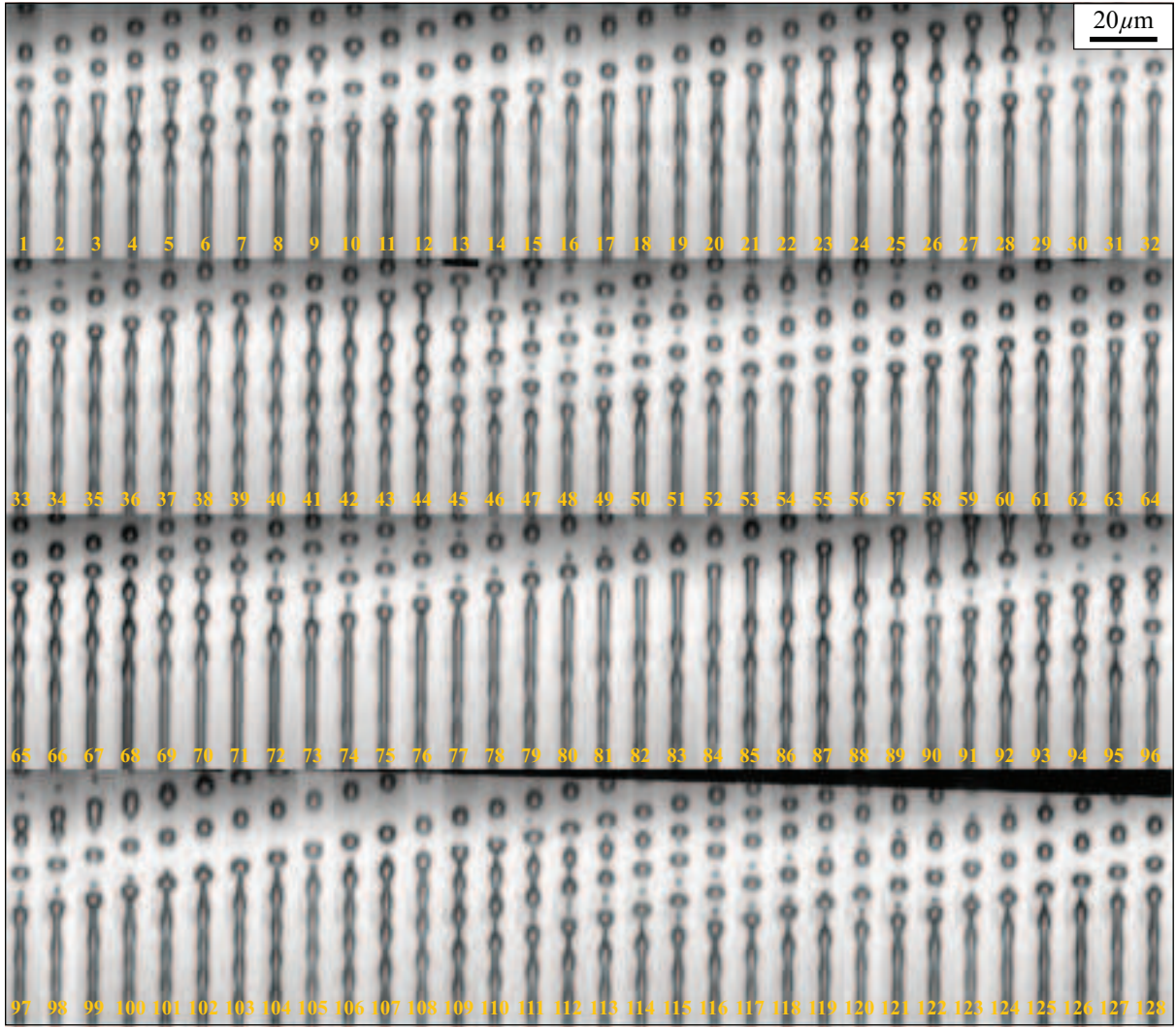}
    \caption{\label{Fig:montage} A sequence of 128 image frames recorded at 13.76\,Mfps showing the breakup of a 1.25\,\micron radius liquid jet into microdroplets. The liquid used was a 0.9\% saline solution, with density $\rho = 1003$\,kg/m$^3$, surface tension $\gamma = 72$\,mN/m, and viscosity $\eta = 1\,\mbox{mPa}\cdot$s. The frame number is indicated in the bottom of each frame. The interframe time is 73\,ns.}
\end{figure*}
The same set of image frames is shown in greater detail in Fig.~\ref{Fig:montage}. It is observed that droplet formation from the breakup of a 1\,\micron liquid microjet shows more irregularities in comparison to droplet formation from a 18.5\,\micron jet (\emph{cf.}~Fig.~\ref{Fig:timeseries}(a)). This is best witnessed by the variation in droplet and satellite droplet spacing and the diversity in droplet pinch-off location.

\begin{figure}
    \includegraphics[]{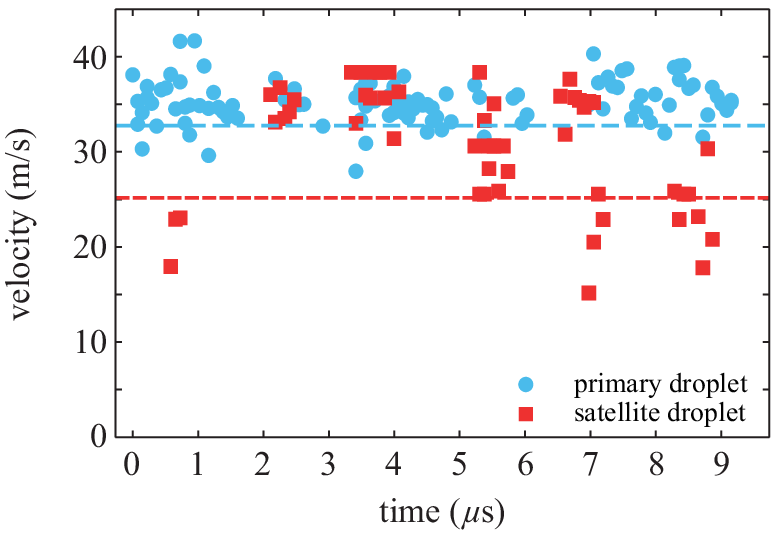}
    \caption{\label{Fig:droplet_velocity_brandaris}Experimental droplet velocity obtained from the 13.76\,Mfps recording of the primary droplet (blue) and the satellite droplet (red) as a function of time. The velocity of the primary droplet $v_\mathrm{prim} = 35\,\pm\,2$\,m/s; the velocity of the satellite droplet $v_\mathrm{sat} = 30\,\pm\,10$\,m/s shows a wide velocity distribution. The predicted velocity by the lubrication approximation model for the primary droplet is $32.8\,\pm\,0.2$\,m/s and for the satellite droplet $25.2\,\pm\,0.1$\,m/s (dashed lines).}
\end{figure}
This is also expressed in the droplet velocity distribution. In Fig.~\ref{Fig:droplet_velocity_brandaris} the droplet and satellite droplet velocity as a function of the time is shown. The velocity of the droplets is obtained by tracking the center of mass of each droplet throughout the high-speed imaging recording using digital image processing. The experimental velocity of the primary droplet and satellite droplet are $35\,\pm\,2$\,m/s and $30\,\pm\,10$\,m/s. The velocity of the satellite droplet shows a wide variation. The velocity predicted by the lubrication approximation model for the primary droplet is 33\,m/s.

\section{\label{Sec:bi}Boundary integral}
\begin{figure}
    \includegraphics[]{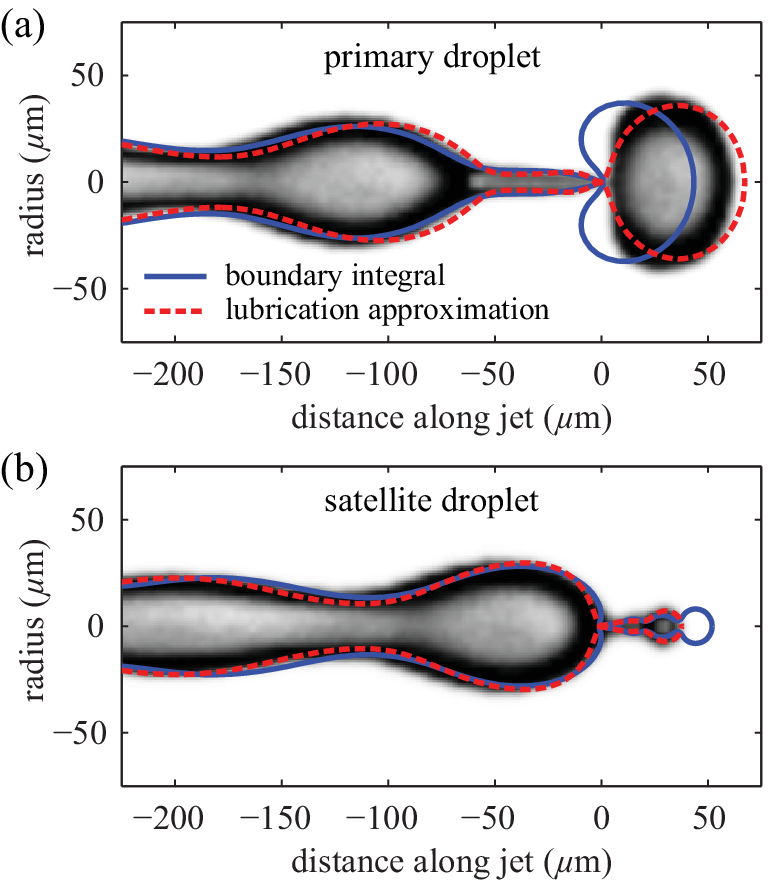}
    \caption{\label{Fig:bi}Comparison between the inviscid, axisymmetric boundary-integral simulation and the one-dimensional lubrication approximation including viscosity. Here, the jet radius is 18.5\,\micron and the jet velocity is 5.4\,m/s. A 40\% glycerol-saline solution was used, with density $\rho_\ell = 1098$\,kg/m$^3$, viscosity $\eta = 3.65\,\mbox{mPa}\cdot$s, and surface tension $\gamma = 67.9$\,mN/m. While the boundary-integral approach can model the overturning of the pinching drop (a), it does not accurately capture the formation of satellites (b).}
\end{figure}

\begin{figure}
    \includegraphics[]{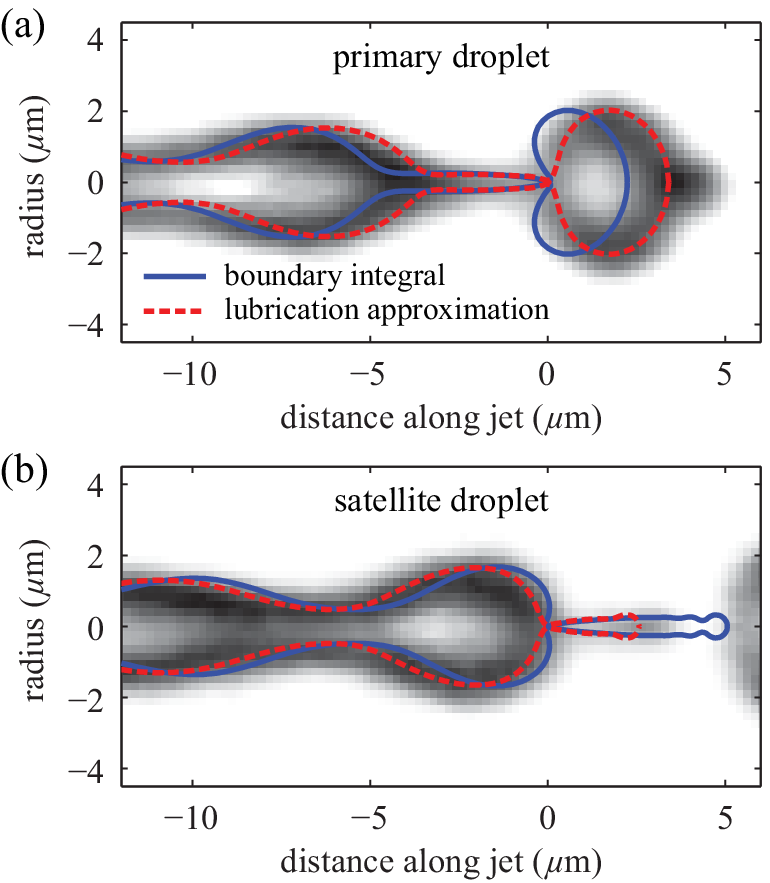}
    \caption{\label{Fig:bi_brandaris}Comparison between the boundary-integral simulation and the one-dimensional lubrication approximation for the breakup of a 1.25\,\micron jet. The jet velocity 35\,m/s. A 0.9\% normal saline solution was used, with density $\rho_\ell = 1003$\,kg/m$^3$, viscosity $\eta = 1\,\mbox{mPa}\cdot$s, and surface tension $\gamma = 72$\,mN/m.}
\end{figure}
The one-dimensional nature of the lubrication approximation model implies that it cannot describe the concave curvature, or `overturning', of the droplet at pinch-off. Droplet overturning typically occurs for low-viscosity liquids, but it is also observed experimentally for more viscous liquids.\cite{Chen2002} The shape of an inviscid liquid droplet at pinch-off exhibits droplet overhang with a unique characteristic angle of 112.8$^\circ$.\cite{Day1998} In Wilkes \emph{et al.}\cite{Wilkes1999} a numerical study is presented on the influence of liquid viscosity on the angle of droplet overhang. Droplet overturning is suppressed when the liquid viscosity is sufficiently large such that viscous shear stress is efficiently dissipated into the liquid.
Droplet overturning is suppressed for $\textrm{Oh}$ typically around 0.01 and 0.1.
In this work we are close to the critical viscosity ($\textrm{Oh} = 0.1$ in both studies) and overturning could occur. However, from the high-speed (shadow) images in Fig.~\ref{Fig:detailed_comparison} and Fig.~\ref{Fig:montage} it is not clear whether the droplet shows some overhang. The amount of droplet overhang, if exists, is of the same order of the pixel size and can not be quantified.

Here we make a comparison between an axisymmetric boundary integral (BI) calculation, which can describe droplet overhang, and the one-dimensional lubrication approximation model (which includes viscosity, in contrast to BI). In Fig.~\ref{Fig:bi} and Fig.~\ref{Fig:bi_brandaris} we show boundary-integral numerical simulations for the same cases as studied above experimentally and within the lubrication approximation.
The drawback is that BI methods are only applicable in the inviscid limit ($\textrm{Oh} = 0$) due to the potential flow description. Details of the implementation of the BI scheme are given in Gekle \emph{et al.}\cite{Gekle2009}

Fig.~\ref{Fig:bi} and Fig.~\ref{Fig:bi_brandaris} shows a comparison of droplet pinch-off obtained from the BI simulation and the lubrication approximation. As can be seen in Fig.~\ref{Fig:bi}(a), the BI simulation indeed displays the overturning of the main droplet just before pinch-off, in contrast to the one-dimensional lubrication theory. However, the size of the predicted overturning is clearly inconsistent with the experimental observations. The experimental results for the droplet is in between the BI and lubrication predictions. In addition, Fig.~\ref{Fig:bi}(b) reveals some deviations in the shape and size of the satellite droplet. We attribute this to the missing viscosity in the BI method as satellite formation is closely connected to viscous effects.

\section{\label{Sec:discussion_conclusion}Discussion \& conclusion}
\begin{figure}
    \includegraphics[]{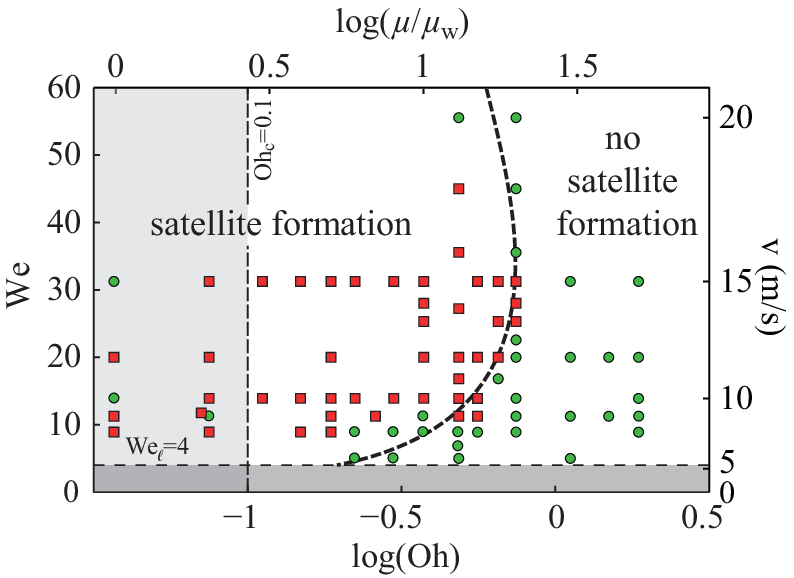}
    \caption{\label{Fig:phasediagram}Phase diagram of satellite droplet formation regimes for various viscosity liquids and jetting velocities. Squares denote cases where our lubrication approximation calculation predicts satellite droplets, circles denote cases without satellite droplets. On the horizontal axis the logarithm of the Ohnesorge number $\log{\textrm{Oh}} = \mu / \sqrt{\rho r \gamma}$ (at the bottom) and the logarithm of the relative viscosity $\log{\mu / \mu_\mathrm{w}}$ (at the top), with $\mu_\mathrm{w} = 1\,\mbox{mPa}\cdot$s the viscosity of water at 20\Celsius, are shown. The vertical axis shows the Weber number $\textrm{We} = \rho v^2 r / \gamma$ (at the left) and the jetting velocity $v$ (at the right). The lower critical velocity for jet formation is indicated by the horizontal dashed line. Droplet `overhang' is suppressed in the region right of the vertical dashed line, with $\textrm{Oh}_\mathrm{c} = 0.1$. The curved dashed line is a guide to the eye. Right of this line the formation of satellite droplets is suppressed. The radius of the jet, liquid density, and surface tension are kept constant ($r = 10$\,\micron, $\rho_\ell = 1000$\,kg/m$^3$, and $\gamma = 72$\,mN/m).}
\end{figure}
In spite of its limitations the one-dimensional lubrication approximation model predicts the formation of the microdroplets and satellite droplets and their sizes with great accuracy. The advantage of a one-dimensional approach is the limited time required to perform a complete calculation. This makes it possible to perform a parameter study. In Fig.~\ref{Fig:phasediagram} a phase diagram is shown, indicating the effect of the jetting velocity and viscosity on the formation of satellite droplets. The jetting velocity is expressed in the Weber number, the viscosity in the Ohnsesorge number. A clear region is found for which the formation of satellite droplets is suppressed.
In this regime the driving frequency and the amplitude of the modulation are kept constant. It is well known that the amplitude and the wavelength-to-diameter ratio affect the formation and behavior of satellite droplets.\cite{Pimbley1977} To get a complete picture the phase diagram should be extended so that it includes these parameters.

In conclusion, the extremely fast droplet formation process from the spontaneous breakup of a liquid microjet is resolved at a high spatial and temporal resolution using ultra high-speed imaging up to 14\,Mfps. A direct comparison is made between the experimental results and those obtained from a one-dimensional (1D) model based on the lubrication approximation. A 1D model is limited by the fact that it can not describe the complex shape of the droplet at the final moment of pinch-off, \emph{e.g.}~droplet `overhang', which typically occurs for low-viscosity liquids. We made a comparison between the lubrication approximation and a boundary integral calculation. The lubrication approximation predicts the shape of the droplet at pinch-off to be closer to its most favorable state -- a perfect sphere -- hence it is less subject to shape oscillations. Inspite of this discrepancy it is shown that the lubrication approximation can predict the size of the droplets, its velocity, and the formation of satellite droplets, with great accuracy.
The 1D origin of the lubrication approximation makes it computationally less demanding in comparison to two-dimensional models, which makes it highly interesting to be used to investigate parameter space in droplet formation, also for diminutive Rayleigh jets.

\acknowledgements
We would like to thank Wietze Nijdam and Jeroen Wissink (\mbox{Medspray} XMEMS bv) for the supply and preparation of the nozzles used in this work. We kindly acknowledge Arjan van der Bos, Theo Driessen, and Roger Jeurissen for very helpful discussions on both theory and experiment. We are grateful to Professor E.~P.~Furlani for discussions on computational aspects. We highly appreciate the skillful technical assistance of Gert-Wim Bruggert, Martin Bos, and Bas Benschop.
This work was financially supported by the MicroNed technology program of the Dutch Ministry of Economic Affairs through its agency SenterNovem under grant \mbox{Bsik-03029.5}.

%\bibliographystyle{unsrt}
%\bibliography{refs}

\begin{thebibliography}{99}
\bibitem{note1}
When scaling all lengths with the jet radius $r$ and time with capillary time
  $\tau_\mathrm{c}$, these equations only contain two dimensionless
  parameters: the Ohnesorge number $\textrm{Oh}$, expressing the importance of
  the viscous term, and the Weber number $\textrm{We}$, expressing the jet
  velocity.
\end{thebibliography}

\end{document}